\documentclass[slac_one]{revtex4}
\usepackage{graphicx}
\usepackage{fancyhdr}
\usepackage{epsf,epsfig}
\def\lappeq{\mathrel{\rlap{\raise.5ex\hbox{$<$}}
{\lower.5ex\hbox{$\sim$}}}}

\def\beq{\begin{equation}}
\def\eeq{\end{equation}}

\def\m12{m_{1\!/2}}

\input paperdef

\begin{document}

\title{{\small{2005 International Linear Collider Workshop - Stanford,
U.S.A.}}\\ 
\vspace{12pt}
Sensitivities to the SUSY Scale 
from Electroweak Precision Observables}

%

\author{J.~Ellis, S.~Heinemeyer}
\affiliation{TH Division, Physics Department, CERN, Geneva, Switzerland}
\author{K.~Olive}
\affiliation{William I.\ Fine Theoretical Physics Institute,
University of Minnesota, Minneapolis, USA}
\author{G.~Weiglein}
\affiliation{Institute for Particle Physics Phenomenology, University of
Durham, UK}

\begin{abstract}
Precision measurements, now and at a future linear electron-positron
collider (ILC), can provide indirect information about the possible
scale of supersymmetry. Performing a $\chi^2$ analysis, we illustrate the 
present-day and possible future ILC
sensitivities within the constrained minimal supersymmetric extension of
the Standard Model (CMSSM), varying the parameters 
so as to obtain the cold dark matter density
allowed by WMAP and other cosmological data.
The current data are in very good agreement with the
CMSSM prediction for $\tb = 10$, with a clear preference for relatively
small values of the universal gaugino mass, $m_{1/2} \sim 300$~GeV. 
In this case, there would be good
prospects for observing supersymmetry directly at both the LHC and
the ILC, and some chance already at the Tevatron collider. 
For $\tb = 50$, the quality of the fit is worse,
and somewhat larger $m_{1/2}$ values are favoured. With the prospective
ILC accuracies the sensitivity to indirect effects of supersymmetry
greatly improves. This may provide indirect access to supersymmetry even
at scales beyond the direct reach of the LHC or the ILC.
\end{abstract}

\maketitle

\thispagestyle{fancy}


\section{INTRODUCTION}

Measurements at low energies may provide interesting indirect information
about the masses of particles that are too heavy to be produced directly.
A prime example is the use of precision electroweak data from LEP, the
SLC, the Tevatron and elsewhere to predict (successfully) the mass of the
top quark and to provide an indication of the possible mass of the
hypothetical Higgs boson~\cite{lepewwg}. 
Predicting the masses of supersymmetric
particles is much more difficult than for the top quark or even the Higgs 
boson, because the renormalizability of the
Standard Model and the decoupling theorem imply that many low-energy
observables are insensitive to heavy sparticles. Nevertheless, present
data on observables such as $\MW$, $\sweff$, $(g-2)_\mu$ and
$\br(b \to s \ga)$ already provide interesting information on the scale of
supersymmetry (SUSY), as we discuss in this paper, and have a 
great potential in view of prospective improvements of experimental and
theoretical accuracies~\cite{ehow3}.

In the future, a linear $e^+ e^-$ collider (ILC) will be the best available
tool for making many precision measurements~\cite{lctdrs}. 
It is important to understand
what information ILC measurements may provide about supersymmetry, 
both for the part of the
spectrum directly accessible at the LHC or the ILC and for sparticles
that would be too heavy to be produced directly.
Comparing the
indirect indications with the direct measurements would be an important
consistency check on the theoretical framework of supersymmetry.

Improved and more complete calculations of the supersymmetric
contributions to a number of low-energy observables such as $\MW$ and
$\sweff$ have recently become available~\cite{PomssmRepetc}. These, 
combined with
estimates of the experimental accuracies attainable at the ILC and
future theoretical uncertainties from unknown higher-order corrections, 
make now an
opportune moment to assess the current sensitivities 
and the projection to the situation at the ILC~\cite{ehow3},
see \citere{prevfits} for previous studies.
In order to reduce the large dimensionality
of even the minimal supersymmetric extension of the
Standard Model, we work here
in the framework of the constrained MSSM (CMSSM), in which the soft
supersymmetry-breaking scalar and gaugino masses are each assumed to be
equal at some GUT input scale. In this case, only four new 
parameters are needed: the universal gaugino mass $m_{1/2}$,
the scalar mass $m_0$, the trilinear soft supersymmetry-breaking parameter
$A_0$, and the ratio $\tb$ of Higgs vacuum expectation values. The 
pseudoscalar Higgs mass $\MA$ and the magnitude of the Higgs mixing 
parameter $\mu$ can be determined by using the electroweak vacuum 
conditions, leaving the sign of $\mu$ as a residual ambiguity.

The non-discovery of supersymmetric particles and the Higgs boson at 
LEP and other present-day colliders imposes significant lower bounds on
$m_{1/2}$ and $m_0$. An important further constraint is provided by the
density of dark matter in the Universe, which is tightly constrained by
WMAP and other astrophysical and cosmological data~\cite{WMAP}. 
These have the effect within the CMSSM, assuming that the dark matter 
consists largely of neutralinos~\cite{EHNOS},
of restricting $m_0$ to very narrow allowed strips
for any specific choice of $A_0$, $\tb$ and the sign of 
$\mu$~\cite{WMAPstrips}. 
Thus, the dimensionality of the supersymmetric parameter space is further
reduced, and one may explore supersymmetric phenomenology along these
`WMAP strips', as has already been done for the direct detection of
supersymmetric particles at the LHC and linear colliders of varying 
energies~\cite{otherAnalyses}.
A full likelihood analysis of the 
CMSSM planes incorporating uncertainties in the cosmological
relic density was performed in \citere{eoss4}.
We extend this analysis to indirect effects of supersymmetry.

We consider the following observables: the $W$~boson mass, $\MW$, the
effective weak mixing angle at the $Z$~boson resonance, $\sweff$, the
anomalous magnetic moment of the muon, \mbox{$(g-2)_\mu$} (we use the 
SM prediction based on the $e^+e^-$ data for the hadronic vacuum
polarization contribution, see the discussion in \citere{ehow3})
and the rare
$b$ decays $\br(b \to s \ga)$ and $\br(B_s \to \mu^+ \mu^-)$, as well
as the mass of 
the lightest $\cp$-even Higgs boson, $\Mh$, and the Higgs branching ratios
$\br(\hbb) / \br(\hWW)$. A detailed analysis of the 
sensitivity of each
observable to indirect effects of supersymmetry, taking into account the
present and prospective future experimental and theoretical
uncertainties, can be found in \citere{ehow3}. We briefly summarize here
the main results on the combined sensitivities at present and for the
ILC and update some of the
results of \citere{ehow3}, which were obtained for $\mt = 178.0 \pm 4.3$~GeV,
using the current experimental central value of $\mt =
172.7 \pm 2.9$~GeV~\cite{mtexpnew}.


\section{PRESENT SITUATION}
\label{sec:combcurr}

We first investigate the combined sensitivity of the four low-energy
observables for which experimental measurements exist at present, namely
$\MW$, $\sweff$, $(g-2)_\mu$ and $\br(b \to s \ga)$. The branching ratio
$\br(B_s \to \mu^+ \mu^-)$, for which at present only an upper 
bound exists, has been discussed separately in \citere{ehow3}. 
We begin with an analysis of the sensitivity to $m_{1/2}$ moving 
along the WMAP
strips with fixed values of $A_0$ and $\tb$. The experimental central
values, the present experimental errors and theoretical uncertainties
are as described in \citere{ehow3}.
The experimental uncertainties, the intrinsic errors from unknown
higher-order corrections and the parametric uncertainties have been
added quadratically, except for $\br(b \to s \ga)$, where they have
been added linearly. Assuming that the four observables are
uncorrelated, a $\chi^2$ fit has been performed with
$\chi^2 \equiv \sum_{n=1}^{N} \KL R_n^{\rm exp} - R_n^{\rm theo} \KR^2/
                                 \KL\si_n \KR^2$.
Here $R_n^{\rm exp}$ denotes the experimental central value of the
$n$th observable, so that $N = 4$ for the set of observables included in
this fit,
$R_n^{\rm theo}$ is the corresponding CMSSM prediction and $\si_n$
denotes the combined error, as specified above.
We have rejected all points of the CMSSM parameter space with either
$\Mh < 111 \gev$~\cite{LEPHiggs,mhiggsAEC} (taking into account
theoretical uncertainties from unknown higher orders)
or a chargino mass lighter than $103 \gev$~\cite{pdg}.

\begin{figure}[thb!]
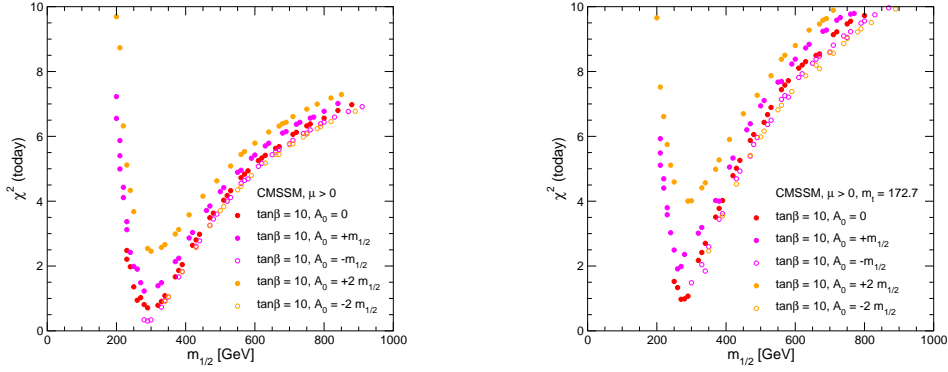

\begin{center}
\epsfig{figure=ehow.CHI11a.cl.eps,width=.30\textwidth}
\hspace{5em}
\epsfig{figure=ehow.CHI11a.1727.cl.eps,width=.30\textwidth}
\caption{%
The results of $\chi^2$ fits based on the current experimental
results for the precision observables $\MW$, $\sweff$, $(g-2)_\mu$ and
$\br(b \to s \ga)$ are shown as functions of $m_{1/2}$ in the CMSSM
parameter space with CDM constraints for $\tb = 10$ and different values of
$A_0$.  
The left plot shows the results for 
$\mt = 178.0 \pm 4.3 \gev$, and the right plot
shows the results for $\mt = 172.7 \pm 2.9 \gev$.
}
\label{fig:CHIoldnew}
\end{center}
\end{figure}

The results for $\tb = 10$ are shown in \reffi{fig:CHIoldnew}
using $\mt = 178 \pm 4.3 \gev$ (left) and 
$\mt = 172.7 \pm 2.9 \gev$~\cite{mtexpnew} (right). 
They indicate that, already at the present level of experimental
accuracies, the electroweak precision observables combined with the WMAP
constraint provide a sensitive probe of the CMSSM, yielding interesting
information about its parameter space. For $\tb = 10$, the CMSSM provides a
very good description of the data, resulting in a remarkably small minimum
$\chi^2$ value. The fit shows a clear preference for relatively small
values of $m_{1/2}$, with a best-fit value of about $m_{1/2} = 300 \gev$.
This minimum is even more pronounced and located at slightly lower
$m_{1/2}$ values for the lower top-quark mass value, where especially
$\MW$ and $\sweff$ favour lower $m_{1/2}$ values.
The best fit is obtained for $A_0 \leq 0$, while positive values of $A_0$
result in a somewhat lower fit quality. 
The fit yields an upper bound on $m_{1/2}$ of about 600 (450)~GeV at the
90\%~C.L.\ (corresponding to $\Delta \chi^2 \le 4.61$)
for $\mt = 178.0 \pm 4.3 \gev$ ($\mt = 172.7 \pm 2.9 \gev$).
Some of the principal contributions to the increase in $\chi^2$ when 
$m_{1/2}$ increases for $\tb = 10$ are as follows (evaluated for 
$\mt = 178.0 \pm 4.3 \gev$).
For $A_0 = - m_{1/2}$, $m_{1/2} = 900$~GeV, we find that $(g - 2)_\mu$ 
contributes about 5 to $\Delta \chi^2$, $\MW$ nearly 1 and $\sweff$
about 0.2, whereas the contribution of $\br(b \to s \gamma)$ is 
negligible. On the other hand, for $A_0 = +2 m_{1/2}$, which is 
disfavoured for $\tb = 10$, the minimum in $\chi^2$ is due to a 
combination of the four observables, but $(g - 2)_\mu$ again gives the 
largest contribution for large $m_{1/2}$. For $\mt = 172.7 \pm 2.9 \gev$ 
the relative contribution of $\MW$ and $\sweff$ to $\Delta \chi^2$
increases, so that also for $A_0 = - m_{1/2}$ the $\Delta \chi^2$ is
more evenly distributed between the observables.

\begin{figure}[htb!]
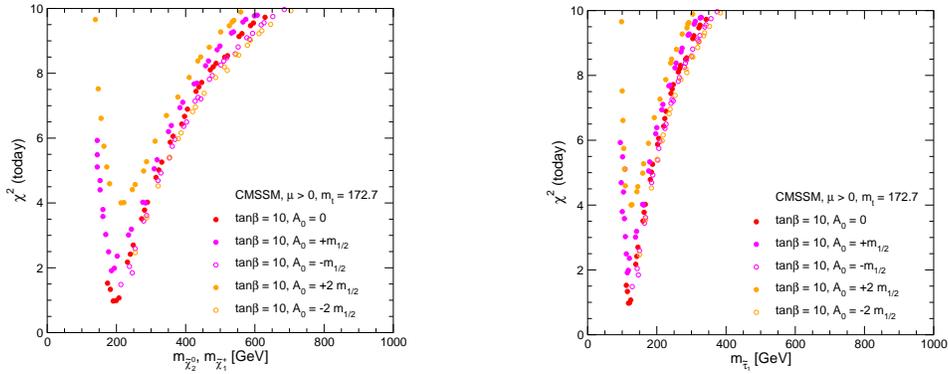

\begin{center}
\epsfig{figure=ehow.mass12a.1727.cl.eps,width=.30\textwidth}
\hspace{5em}
\epsfig{figure=ehow.mass17a.1727.cl.eps,width=.30\textwidth}
\caption{%
The $\chi^2$ contours in the CMSSM with $\tb = 10$ for 
the mass of the lighter chargino (left),
$m_{\chone} \approx m_{\netwo}$, and the 
lighter stau (right), $m_{\tilde{\tau}_1}$, 
based on the fits to the parameter space shown in the right plot of
\reffi{fig:CHIoldnew} (updated from \citere{ehow3} for 
$\mt = 172.7 \pm 2.9 \gev$).
}
\label{fig:massa}
\end{center}
\end{figure}

In \reffi{fig:massa} the fit results of \reffi{fig:CHIoldnew} (right
plot, $\mt = 172.7 \pm 2.9 \gev$) for $\tb = 10$ are expressed in terms of the
masses of the lighter chargino, $m_{\chone} \approx m_{\netwo}$, and
the mass of the lighter stau, $m_{\tilde{\tau}_1}$ 
(updated from \citere{ehow3}).
The best-fit value for $m_{\chone}$
is about 200~GeV, while for $m_{\tilde{\tau}_1}$ the best-fit value is
even below 150~GeV.
The best-fit values for the masses of the neutralinos, charginos and
sleptons~\cite{ehow3} offer good
prospects of direct sparticle detection at both the ILC~\cite{lctdrs}
and the LHC~\cite{lhctdrs}, allowing a detailed determination of their
properties~\cite{lhclc}.
There are also some prospects for detecting the associated 
production of charginos and neutralinos at the Tevatron collider, via 
their trilepton decay signature, in particular. This is estimated to be 
sensitive to $m_{1/2} \lappeq 250$~GeV~\cite{BH}, covering much of the 
region below the best-fit value of $m_{1/2}$ that we find for 
$\tan \beta = 10$.

\begin{figure}[thb!]
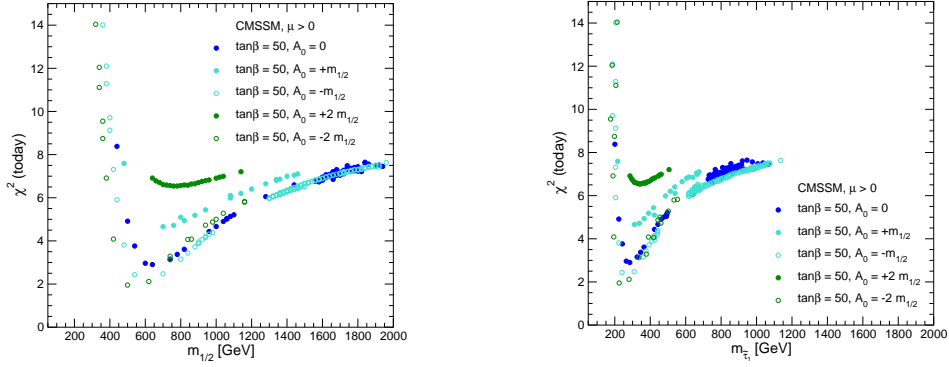

\begin{center}
\epsfig{figure=ehow.CHI11b.cl.eps,width=.30\textwidth}
\hspace{5em}
\epsfig{figure=ehow.mass17b.cl.eps,width=.30\textwidth}
\caption{%
The results of a $\chi^2$ fit based on the current experimental
results for the precision observables $\MW$, $\sweff$, $(g-2)_\mu$ and
$\br(b \to s \ga)$ are shown as functions of $m_{1/2}$ in the CMSSM
parameter space with CDM constraints for different values of
$A_0$, $\mt = 178.0 \pm 4.3 \gev$ and $\tb = 50$ (left).
The corresponding $\chi^2$ contours for the mass of the 
lighter stau, $m_{\tilde{\tau}_1}$, are shown in the right plot.
}
\label{fig:tb50}
\end{center}
\end{figure}

For $\tb = 50$ the overall fit quality is worse than for $\tb = 10$, and
the sensitivity to $m_{1/2}$ from the precision observables is lower,
as is shown in \reffi{fig:tb50} (left). This
is related to the fact that, whereas $\MW$ and $\sweff$ prefer small values
of $m_{1/2}$ also for $\tb = 50$, 
the CMSSM predictions for $(g-2)_{\mu}$ and $\br(b \to s \gamma)$ for high 
$\tb$ are in better agreement with the data for larger
$m_{1/2}$ values, 
see the discussion in \citere{ehow3}.
Also in
this case the best fit is obtained for negative values of $A_0$, but the
preferred values for $m_{1/2}$ are 200--300~GeV higher than for $\tb = 10$.

In the case $\tb = 50$ (for $\mt = 178 \pm 4.3 \gev$) the best-fit 
values for the LSP mass and the lighter stau are still below about
250~GeV~\cite{ehow3}. The fit result for the stau mass is shown in the right
plot of \reffi{fig:tb50}. 
The minimum $\chi^2$ for the other masses is shifted upwards compared to
the case with $\tb = 10$. The best-fit values are obtained 
in the region
400--600~GeV. Correspondingly, these sparticles would be harder to
detect. At the ILC with $\sqrt{s} \lsim 1 \tev$, the best prospects would 
be for the production of $\neu{1}\neu{2}$ or of $\Staue\AStaue$. Other
particles can only be produced if they turn out to be on the light
side of the $\chi^2$~function.

For the case of $\tb = 10$, all the coloured particles should 
be accessible at the LHC~\cite{ehow3}. However, among them, only
${\tilde t}_1$ has a substantial part of its $\chi^2$-favoured spectrum 
below $500 \gev$, which would allow its detection at the ILC. The same
applies for the mass of the $A$~boson. The Tevatron collider has a 
sensitivity to $\mste \lappeq 450$~GeV~\cite{BH}, which is about
our best-fit value for $\tan \beta = 10$.
For the $\tb = 50$ case (with $\mt = 178.0 \pm 4.3$~GeV)
the particles are mostly inaccessible at the ILC, 
though the
LHC has good prospects. However, at the 90\%~C.L.\ the coloured sparticle 
masses
might even exceed $\sim 3 \tev$, which would render their detection difficult.
Concerning the heavy Higgs bosons, their masses may well be below $\sim 1
\tev$. In the case of large $\tb$, this might allow their detection via
the process $b \bar b \to b \bar b H/A \to b \bar b \;
\tau^+\tau^-$~\cite{heavyHiggsLHC}. 

\begin{figure}[htb!]
\begin{center}
\epsfig{figure=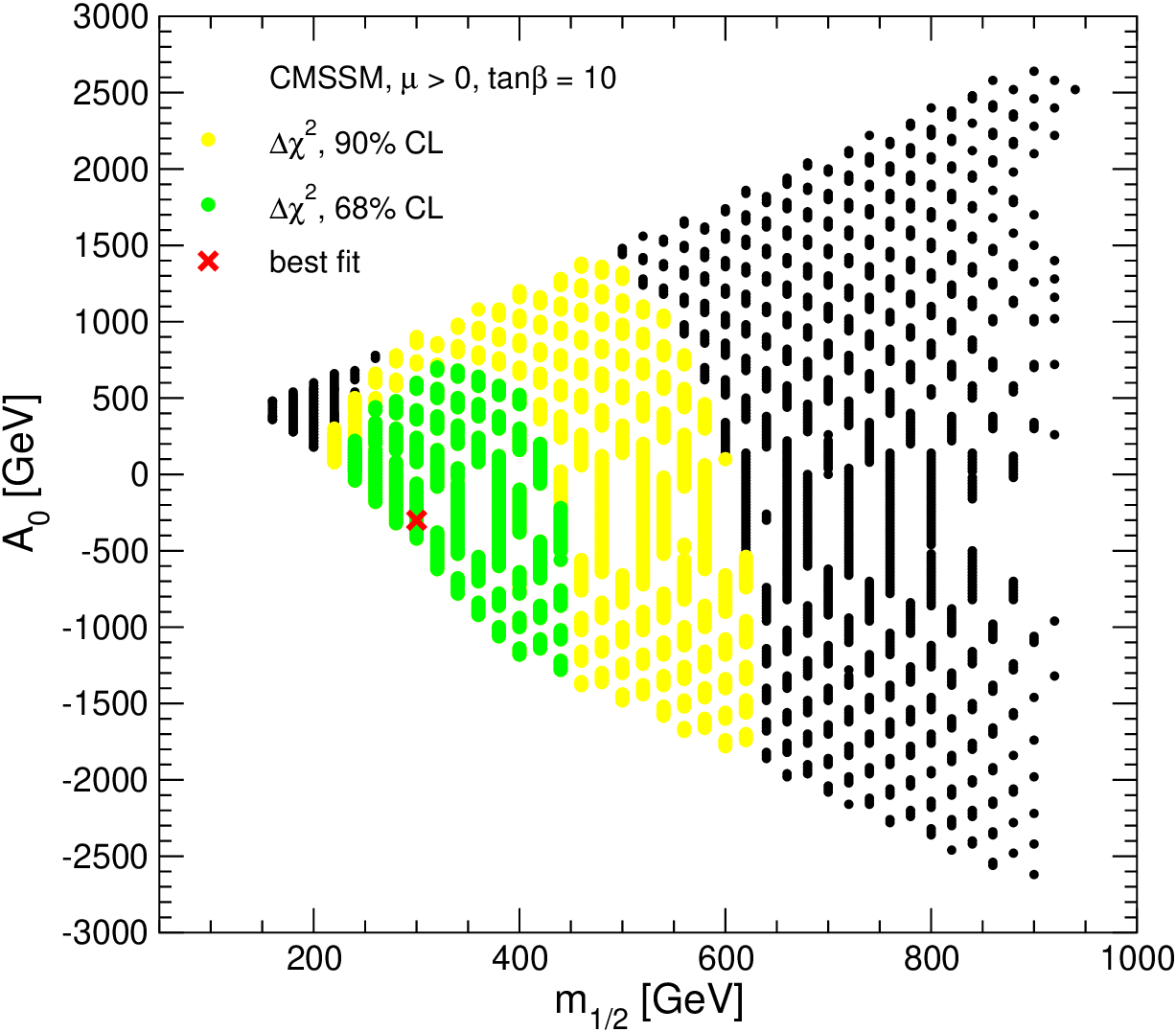,width=.30\textwidth,height=16em}
\hspace{5em}
\epsfig{figure=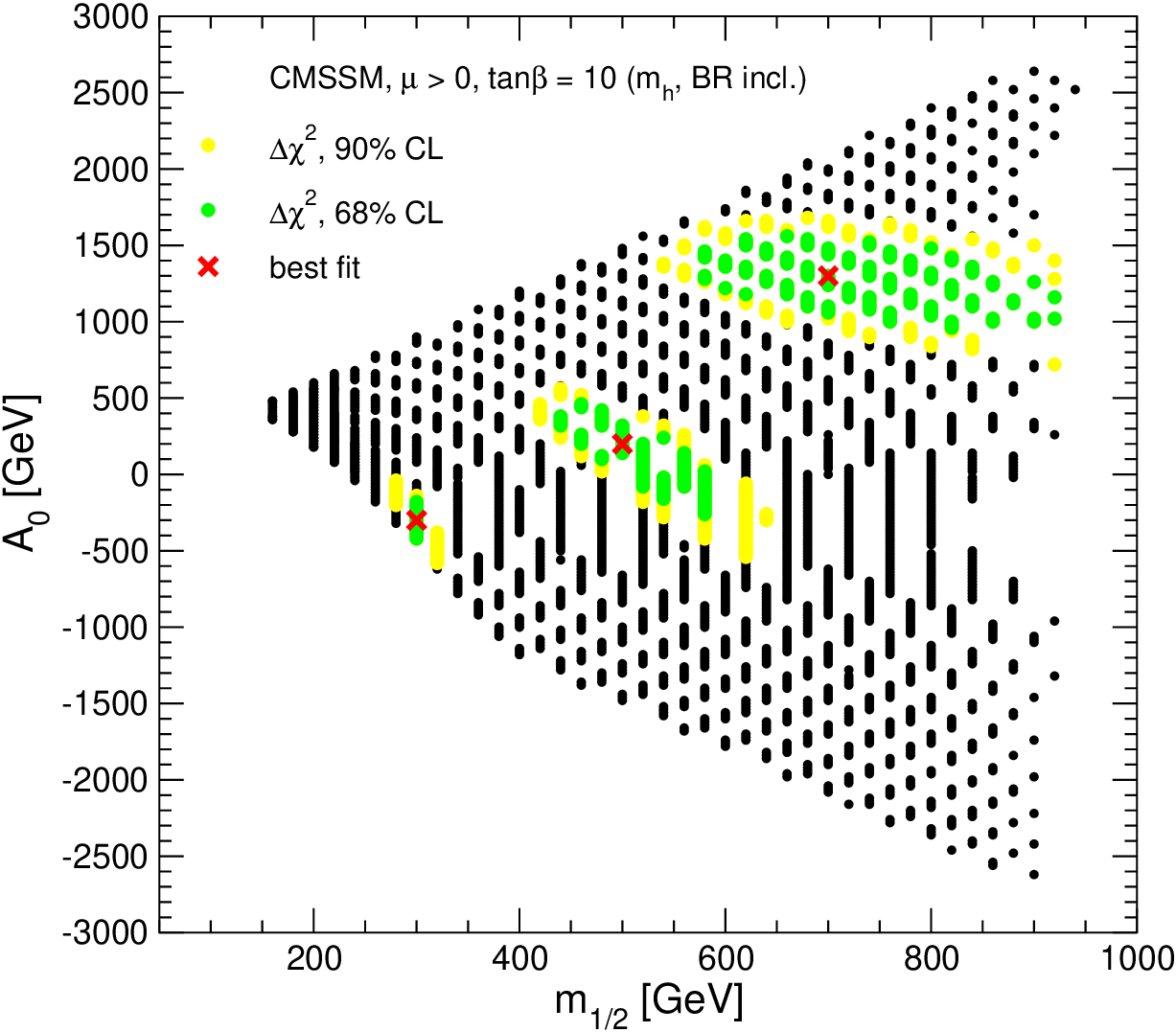,width=.30\textwidth,height=16em}
\caption{%
The results of $\chi^2$ fits for $\tb = 10$
based on the current experimental results (left) and for the
anticipated ILC precision (right) for  
$\MW$, $\sweff$, $(g-2)_\mu$ and
$\br(b \to s \ga)$ (including also $\Mh$ and $\br(\hbb) / \br(\hWW)$
in the right plot) are shown in the \plane{m_{1/2}}{A_0}s of the CMSSM
with the WMAP constraint. The current best-fit point is indicated in
both plots. In the right plot two further hypothetical future
`best-fit' values are shown for illustration.
The coloured regions correspond to the 68\% and 90\% C.L.\ regions,
respectively. 
}
\label{fig:scancurrent}
\end{center}

\vspace{-1em}

\end{figure}

While in the fits presented above we kept 
$A_0/m_{1/2}$ fixed, we 
now analyse the combined sensitivity of the precision observables $\MW$,
$\sweff$, $\br(b \to s \ga)$ and $(g-2)_\mu$ in a scan over the $(m_{1/2},
A_0)$ parameter plane. In order to perform this scan, we have evaluated 
the
observables for a finite grid in the ($m_{1/2}$, $A_0$, $m_0$) parameter
space, fixing $m_0$ using the WMAP constraint. We restrict ourselves
here to the $\tb = 10$ case with $\mt = 178 \pm 4.3 \gev$.
\reffi{fig:scancurrent} (left) shows the WMAP-allowed regions in the
\plane{m_{1/2}}{A_0}.
The current best-fit value
obtained via $\chi^2$ fit is indicated.%
\footnote{A similar analysis within the CMSSM as the one presented here
has recently been carried out in
\citere{allanachlester}, where a fit has been performed involving all
CMSSM parameters. In this analysis, which did not take into account the
electroweak precision observables $\MW$ and $\sweff$, a preference for
larger values of $\tb$ and somewhat larger $m_{1/2}$ has been reported.}
The coloured regions around the best-fit value correspond to the 68\% and 
90\% C.L.\ regions (corresponding to $\Delta \chi^2 \le 2.30, 4.61$,
respectively). 
The precision data yield sensitive constraints on the available
parameter space for $m_{1/2}$ within the WMAP-allowed region. The
precision data are less sensitive to $A_0$.
The 90\% C.L.\ region contains all the WMAP-allowed $A_0$ values in
this region of $m_{1/2}$ values. As expected from the discussion above,
the best fit is obtained for negative $A_0$ and relatively small values
of $m_{1/2}$. At the 68\% C.L., the fit
yields an upper bound on $m_{1/2}$ of about 450~GeV. This bound is
weakened to about 600~GeV at the 90\% C.L.
As discussed above, the overall fit quality is worse for $\tb = 50$, and
the sensitivity to $m_{1/2}$ is less pronounced (plot not shown here). 
In this case the best fit is obtained for
$m_{1/2} \approx 500 \gev$ and negative $A_0$. The upper bound on $m_{1/2}$
increases to nearly 1~TeV at the 68\% C.L.%

\section{ILC PRECISION}
\label{sec:combfuture}

We now turn to the analysis of the future sensitivities of the precision
observables, based on the prospective experimental accuracies at the ILC
and the estimates of future theoretical uncertainties discussed in
\citere{ehow3}. 
We perform a $\chi^2$ fit for the combined sensitivity of the
observables $\MW$, $\sweff$, $(g-2)_\mu$, $\br(b \to s \ga)$, $\Mh$ and
$\br(\hbb) / \br(\hWW)$ in the \plane{m_{1/2}}{A_0} of the CMSSM
assuming ILC accuracies.
The right plot of \reffi{fig:scancurrent} shows the corresponding fit
results for $\tb = 10$. The WMAP-allowed region and the
best-fit point according to the current situation (see the left plot of
\reffi{fig:scancurrent}) are indicated. Two further
hypothetical future `best-fit' points have been chosen for illustration.
For all the `best-fit' points, the assumed central experimental values of
the observables have been chosen such that they precisely coincide with the
`best-fit' points%
\footnote{
We have checked explicitly that assuming future experimental values of the 
observables with values distributed statistically around the present 
`best-fit' points with the estimated future errors does not degrade 
significantly the qualities of the fits.
}%
. The coloured regions correspond to the 68\% and 90\%
C.L.\ regions around each of the `best-fit' points according
to the ILC accuracies.

The comparison of \reffi{fig:scancurrent} (right) with the result
of the current fit (left) shows that the ILC
experimental precision will lead to a drastic improvement in the
sensitivity to $m_{1/2}$ and $A_0$ when confronting precision data with the
CMSSM predictions. 
The comparison of these indirect predictions for $m_{1/2}$ and $A_0$
with the information from the direct detection of supersymmetric particles
would provide a stringent test of the CMSSM framework at the loop level. A
discrepancy could indicate that supersymmetry is realised in a more
complicated way than is assumed in the CMSSM.
For the best-fit values of the current fit 
the ILC precision would allow one to narrow down the
allowed CMSSM parameter space to very small regions in the 
\plane{m_{1/2}}{A_0}. For the
example shown here with best-fit values around $m_{1/2} = 300 \gev$
it is possible to constrain particle masses within about $\pm 10\%$
at the 95\% C.L.\ from the comparison of the precision data with the
theory predictions.
Because of the decoupling
property of supersymmetric theories, the indirect constraints become
weaker for increasing $m_{1/2}$. 
The additional hypothetical `best-fit' points shown in
\reffi{fig:scancurrent} (right) illustrate the indirect
sensitivity to the CMSSM parameters in scenarios where the precision
observables prefer larger values of $m_{1/2}$. We find that for a
`best-fit' value of $m_{1/2}$
as large as 1~TeV, which would lie close to the LHC limit and beyond the
direct-detection reach of the
ILC, the precision data would still allow one to establish an upper
bound on $m_{1/2}$ within the WMAP-allowed region. Thus, this indirect
sensitivity
to $m_{1/2}$ could give important hints for supersymmetry searches at
higher-energy colliders. 
For `best-fit' values of $m_{1/2}$ in excess of
1.5~TeV, on the other hand, the indirect effects of heavy sparticles
become
so small that they are difficult to resolve even with ILC
accuracies~\cite{ehow3}.
Whilst the present analysis has been restricted to the CMSSM, similar 
conclusions
are expected to apply also in models beyond the CMSSM. Such an analysis
is currently in preparation.




\end{document}